\documentclass{kapproc} 

\usepackage{t1enc}

\usepackage{procps} 

\usepackage[dvips]{graphicx}
\usepackage[dvips]{epsfig}

\upperandlowercase

\kluwerbib 

\begin{document}

\articletitle{A Formation Scenario for the
      Heavy-Weight Cluster W3 in NGC 7252}

\chaptitlerunninghead{Formation of W3}

\author{M. Fellhauer and P. Kroupa}
\affil{Sternwarte University Bonn, Germany}
\email{mike,pavel@astro.uni-bonn.de}

\anxx{M.Fellhauer and P.Kroupa}

\begin{abstract}
  The young star cluster W3 (age $300-500$~Myr) in
  NGC~7252 is the most luminous star cluster known to date.  Dynamical 
  mass estimates result in $8 \pm 2 \cdot 10^{7}$~M$_{\odot}$.  With an
  effective radius of about $18$ parsec and a velocity dispersion of
  $45$~kms$^{-1}$ this object is rather one of the recently discovered  
  ultra-compact dwarf (UCD) galaxies than a star cluster.
  
  In our model we propose a formation scenario for W3.
  Observations of interacting galaxies reveal regions of strong star
  formation forming dozens up to hundreds of star clusters in confined
  regions of several hundred parsec in diameter.  The total mass of new
  stars in these regions can reach $10^{7}$ or even $10^{8}$ solar
  masses.  By means of numerical simulations we have shown that the
  star clusters in these regions merge on short time-scales forming
  objects like UCD galaxies or W3.
\end{abstract}


\section{Introduction}
In the merger remnant system NGC~7252 Maraston et al.\ (2004) found
the most luminous star cluster known to date, W3. They determined its
mass via two independent ways.  First they used the total luminosity
of $M_{V} = -16.2$ and a stellar $M/L$ derived from a single stellar
population model to estimate its mass to be $7.2 \cdot
10^{7}$~M$_{\odot}$.  As an independent approach they calculated the
dynamical mass by measuring the velocity dispersion, which turned out
to be quite high, $\sigma = 45 \pm 5$~kms$^{-1}$, and the effective
radius of W3 $R_{\rm eff} = 17.5 \pm 1.8$~pc.  This translates to a
mass of W3 of $M_{\rm dyn} = 8 \pm 2 \cdot  10^{7}$~M$_{\odot}$
assuming dynamical equilibrium of a spherical object with isotropic
velocity distribution.  The age of this object is about
$300$--$500$~Myr (Maraston et al.\ 2004 and references therein), which
indicates that it probably formed during the merger event of the
host system.  The projected distance of W3 to the centre of NGC~7252
is about $10$~kpc and W3 appears to lie within the optical
radius of the galaxy.  The size and mass of this object leads to the
suggestion that it may be one of the recently discovered ultra compact
dwarf galaxies (UCD) found in the Fornax cluster (Hilker et al.\ 1999;
Phillipps et al.\ 2000), rather than an 'ordinary' globular cluster. 
The UCD galaxies have been suggested to be the cores of stripped
nucleated dwarf galaxies (Bekki et al.\ 2003, Mieske et al.\ 2004).

We propose instead a formation scenario which is closely related to
the massive star-bursts caused by the interaction of two gas-rich disc
galaxies.  In interacting systems like the Antennae (NGC~4038/39;
Whitmore et al.\ 1999, Zhang \& Fall 1999) regions of very intense
star-formation arise as a result of the tightly compressed
interstellar media.  Dozens and up to hundreds of young massive star
clusters are observed to form in star cluster complexes (or {\it
  super-clusters}) spanning up to a few hundred pc in diameter.
Kroupa (1998) argues that these super-clusters have to be bound
objects because their age ($\approx 10$~Myr) indicates that they
should be already dispersed.  Simulating super-clusters, by means of
stellar dynamical $N$-body simulations, Fellhauer et al.\ (2002) found
that the star clusters within these super-clusters merge on very short
time-scales (a few dozens to a few hundred Myr), namely a few
crossing-times of the super-cluster.

The age-estimate of W3 points to the possibility that it may have
formed during the merger event of the host-galaxy (i.e.\ out of
merging star clusters) rather than it being the stripped core of a
dwarf galaxy, which should be very old.

In this project we perform stellar dynamical N-body simulations to
show that the merging of star clusters in dense star cluster complexes
is able to form massive objects like W3 in NGC~7252.

\section{Setup}

The simulations are carried out with the particle-mesh code {\sc
  Superbox} with high-resolution sub-grids which stay focussed on the
simulated objects (Fellhauer et al.\ 2000).  

In our models the super-cluster is initially represented by a Plummer
sphere with a Plummer radius of $100$~pc and a cut-off radius of
$500$~pc.  The clusters inside this super-cluster have a total mass of
$9.9 \cdot 10^{7}$~M$_{\odot}$, which leads to a crossing time of the
super-cluster of $9.3$~Myr.  Inside the Plummer sphere of the 
super-cluster, the 'particles' have positions and velocities according   
to the Plummer distribution function.  The 'particles' themselves are
Plummer spheres representing the star clusters with, respectively,
Plummer radii of $4$ or $10$~pc, masses of $10^{6}$ and $5 \cdot
10^{6}$~M$_{\odot}$ and crossing times of $0.75$ and $1.32$~Myr, being 
represented by $10^{5}$ and $5 \cdot 10^{5}$~particles.  Altogether
the super-cluster is filled with $75$ star clusters, from which $69$
are light ones and $6$ are of the heavy type to mimic a mass spectrum
similar to the one found in the young massive star clusters in the
Antennae (Zhang \& Fall 1999).  

This super-cluster is now placed into an analytical galactic potential
consisting of a logarithmic potential for the underlying halo, a
Plummer-Kuzmin disc and a Hernquist bulge, which add up to an almost
flat rotation curve of $220$~kms$^{-1}$.  The super-cluster is placed
at a distance of $20$~kpc initially (apogalacticon) on an eccentric
orbit with perigalacticon of $10$~kpc.

\section{Results}

\begin{figure}[t]
  \centering
  \epsfxsize=03.2cm
  \epsfysize=03.2cm
  \epsffile{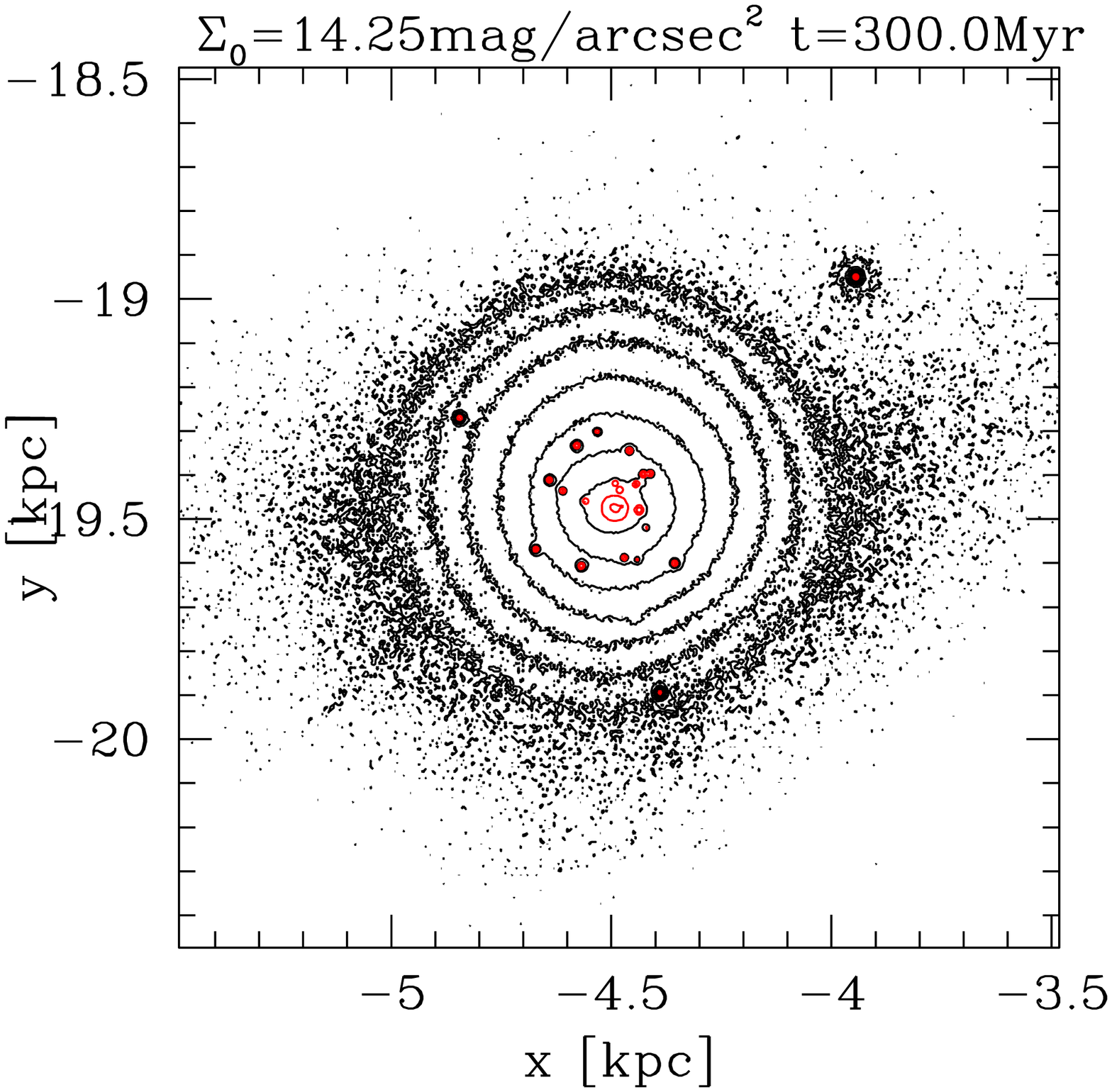}
  \epsfxsize=03.2cm
  \epsfysize=03.2cm
  \epsffile{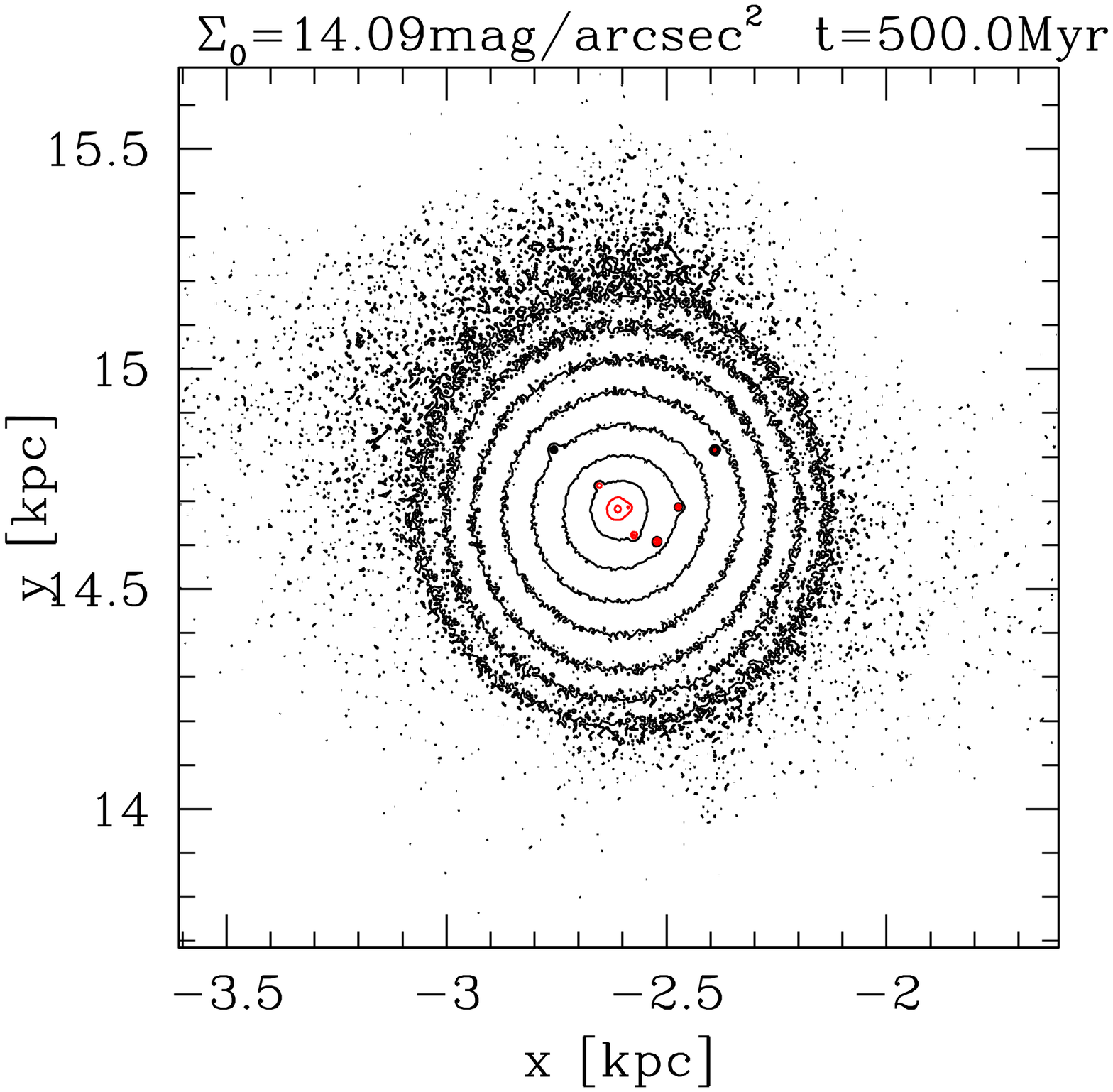}
  \epsfxsize=03.2cm
  \epsfysize=03.2cm
  \epsffile{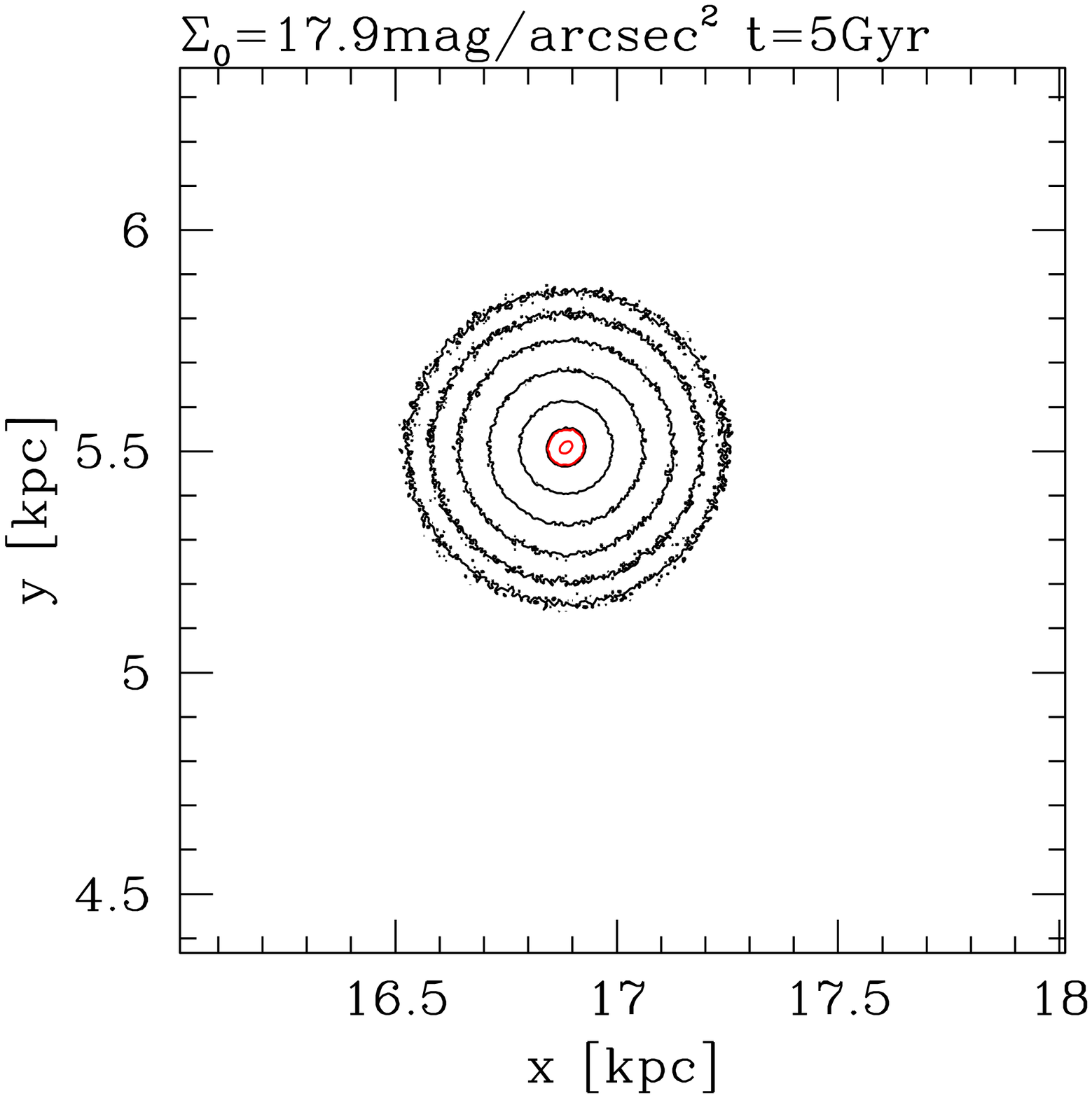}
  \epsfxsize=03.2cm
  \epsfysize=03.2cm
  \epsffile{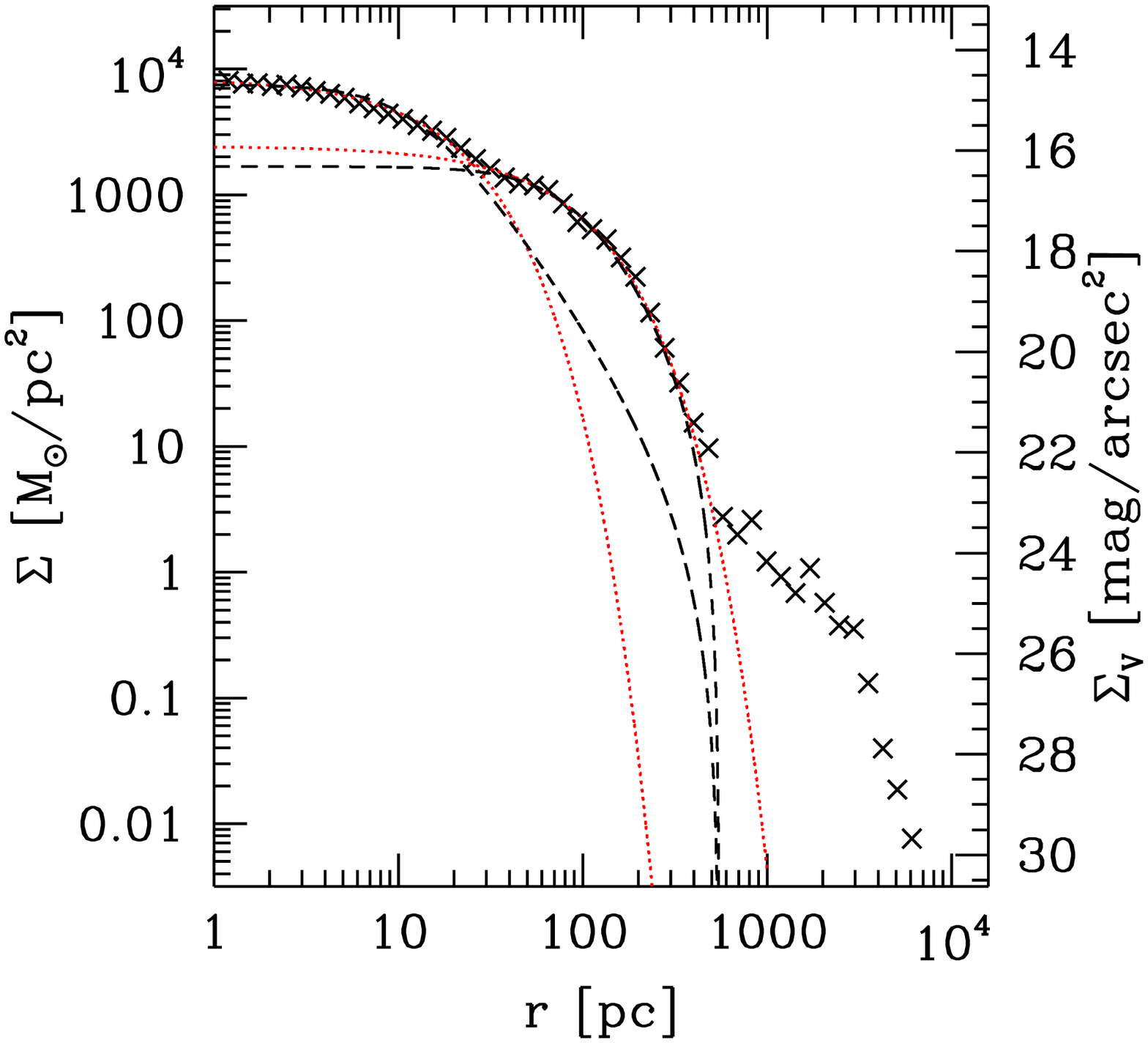}
  \epsfxsize=03.2cm
  \epsfysize=03.2cm
  \epsffile{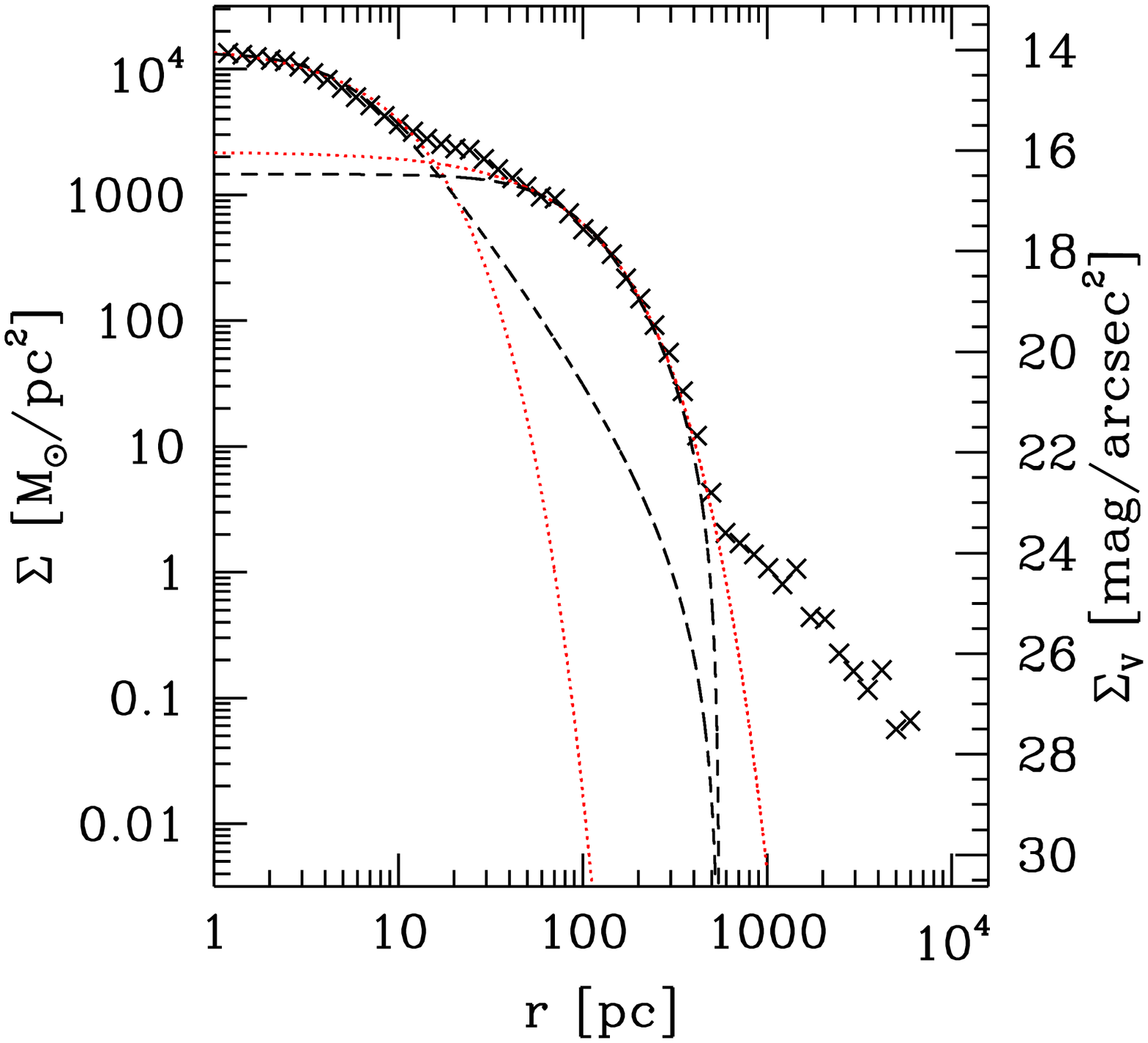}
  \epsfxsize=03.2cm
  \epsfysize=03.2cm
  \epsffile{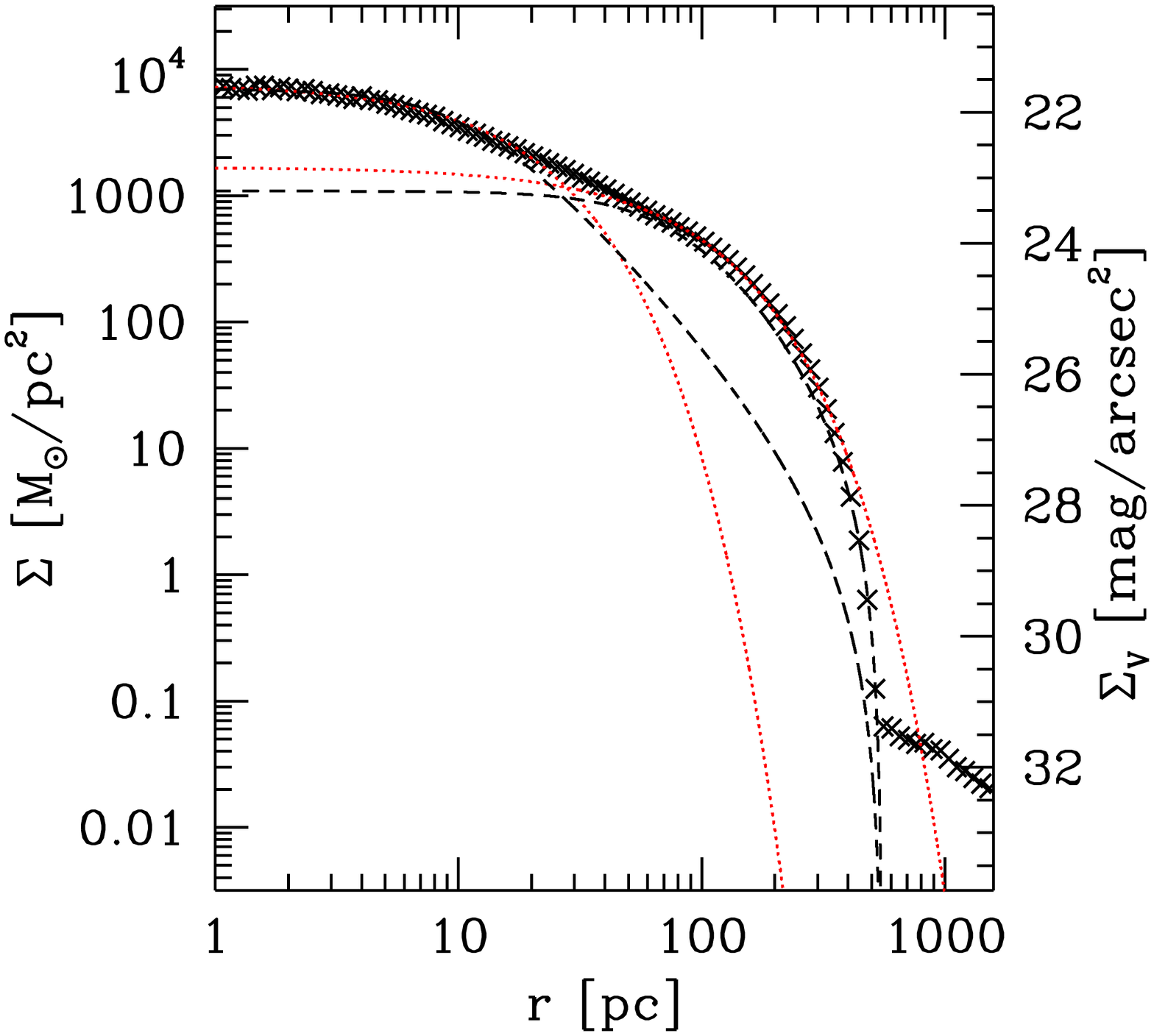}
  \epsfxsize=03.2cm
  \epsfysize=03.2cm
  \epsffile{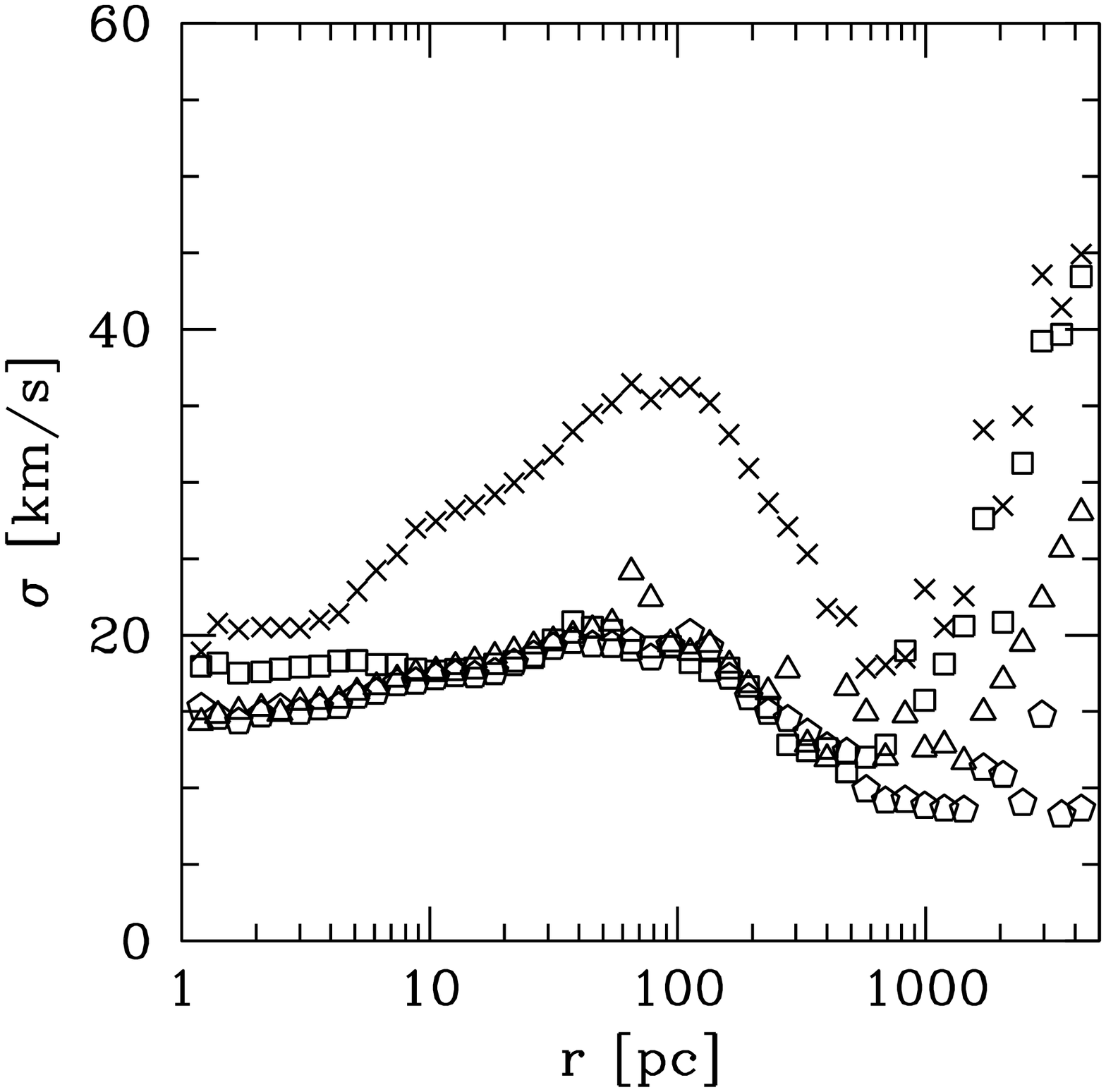}
  \epsfxsize=03.2cm
  \epsfysize=03.2cm
  \epsffile{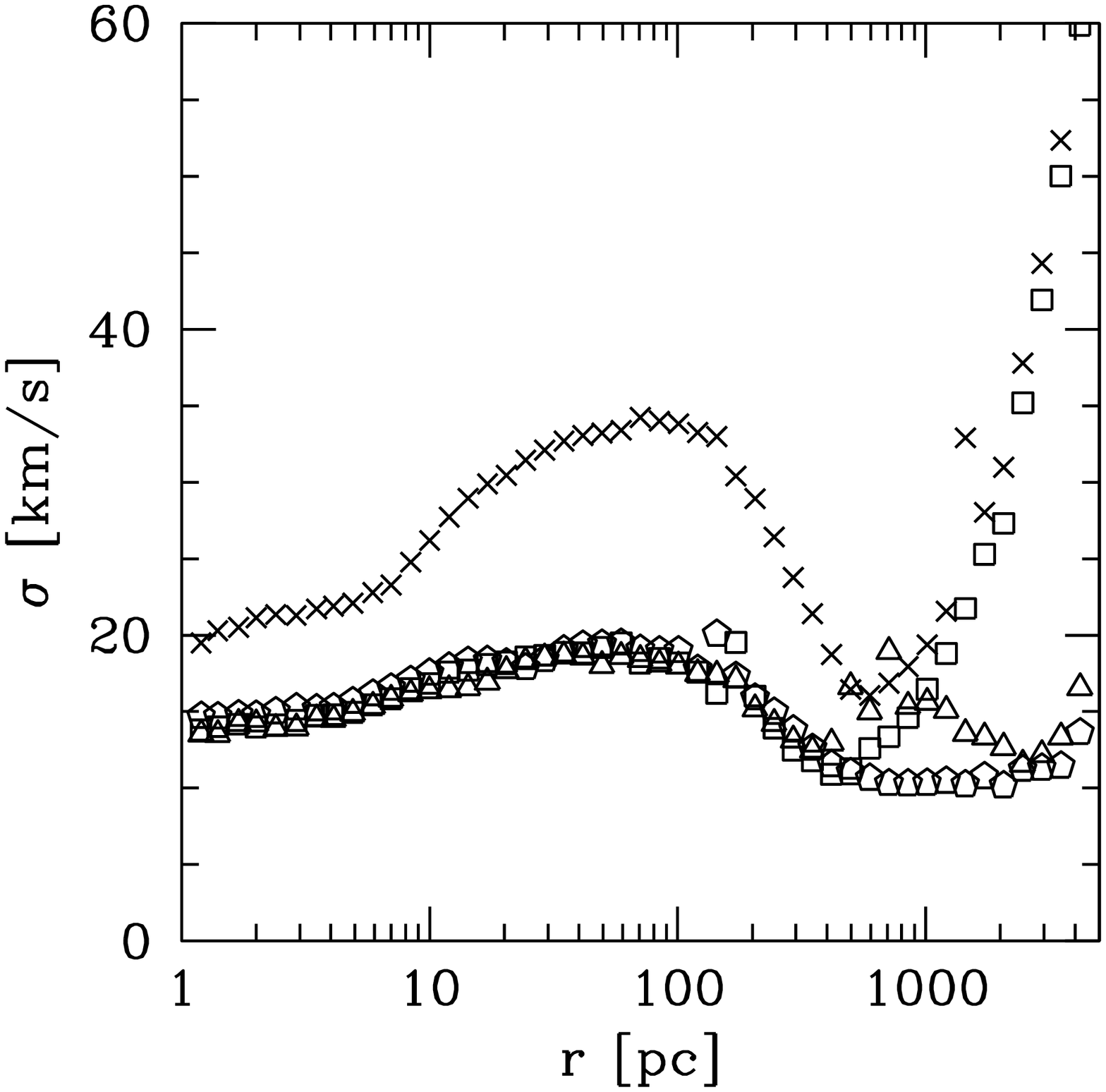}
  \epsfxsize=03.2cm
  \epsfysize=03.2cm
  \epsffile{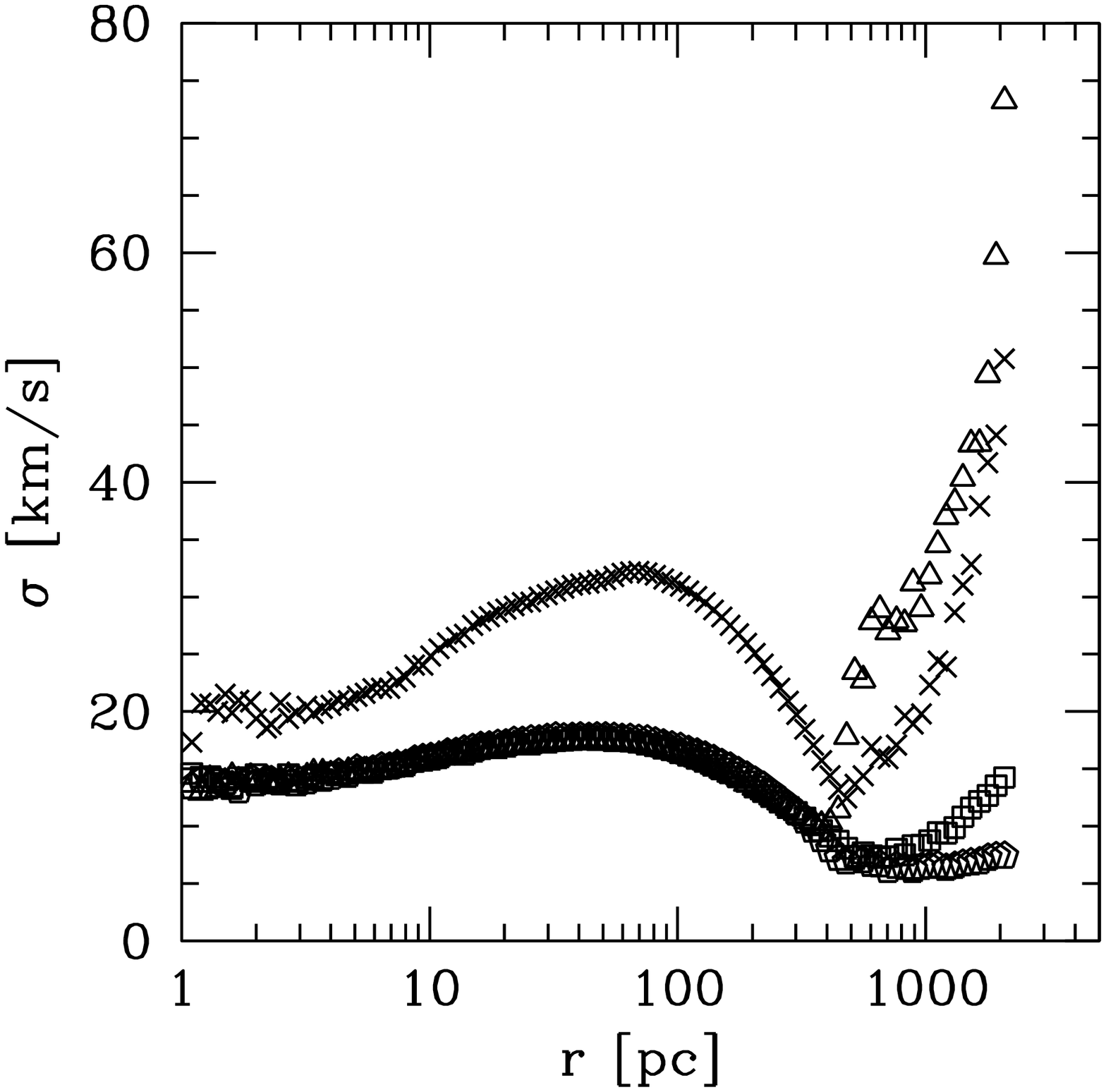}
  \caption{Left: W3 model at $t=300$~Myr; Middle: W3 model at
  $t=500$~Myr.  Right: Merger Object at $5$~Gyr.  
  Top: Contour plot taking a $M/L=0.15$ (right panel $M/L=3.0$).
  Contours have magnitude spacings; outermost contour corresponds to a
  surface brightness of $23$~mag.arcsec$^{-2}$ (righ panel: $25$).  
  Middle: Surface density
  profile of the merger object.  Crosses are the data points, dashed
  lines are the fit with two King profiles, dotted lines are the fit
  with two exponentials.  Brightnesses on the right are calculated
  using $M/L=0.15$ ($3.0$).  
  Bottom: Velocity dispersions; open symbols show
  the line-of-sight dispersions measured in concentric rings around
  the centre of density of the object along the coordinate axes.
  Crosses are the 3D velocity dispersion measured in concentric
  shells.}  
  \label{fig:t300}
\vspace*{-0.5cm}
\end{figure}
 
In contrast to our previous models (Fellhauer \& Kroupa 2002a,b) where
the merger objects were smaller and not that heavy, the merging
process here takes more time.  After the first clusters have merged
and build up a massive and extended object with the remaining clusters
within, it is not encounters between clusters and the merger object
which govern the merging process (the clusters are already within the
object) but dynamical friction acting on the remaining clusters.
Nevertheless a massive merger object is present from an early stage on.

The top row of Figure~\ref{fig:t300} shows the contour plot of the
merger object (left at $t=300$ and right at $t=500$~Myr comprising the
suggested age-range of W3).  The contours are spaced in magnitude
intervals and masses are converted to luminosities taking a
mass-to-light ratio of 0.15.  According to a single stellar population
model run with Starburst99 (Leitherer et al.\ 1999) a stellar population 
of this age has a $M/L=0.1$--$0.2$.  In our model these two
time-slices mark the transition between the time, when still many star
clusters are visible as separate entities and the time when most of
the star clusters have already merged or are dissolved within the
merger object.  

Another interesting feature of our model is that the merger object
shows a cuspy structure with a dynamically cold core.  The
surface-density profile can 
be fitted by either two King profiles or two exponentials
(Fig.~\ref{fig:t300}, middle row) and the velocity dispersion is
rising in the innermost part and reaches its maximum beyond the
transition between the core and the envelope (Fig.~\ref{fig:t300},
bottom row).  The outer part of the velocity dispersion, reaching from
the maximum value to the tidal radius, can be fitted with an
exponential profile with an exponential scale length of the order of
the tidal radius of the object.  Beyond the tidal
radius the velocity dispersion is rising again due to the large
velocities of the extra tidal stars. 

\begin{figure}[t]
  \centering
  \epsfxsize=03.2cm
  \epsfysize=03.2cm
  \epsffile{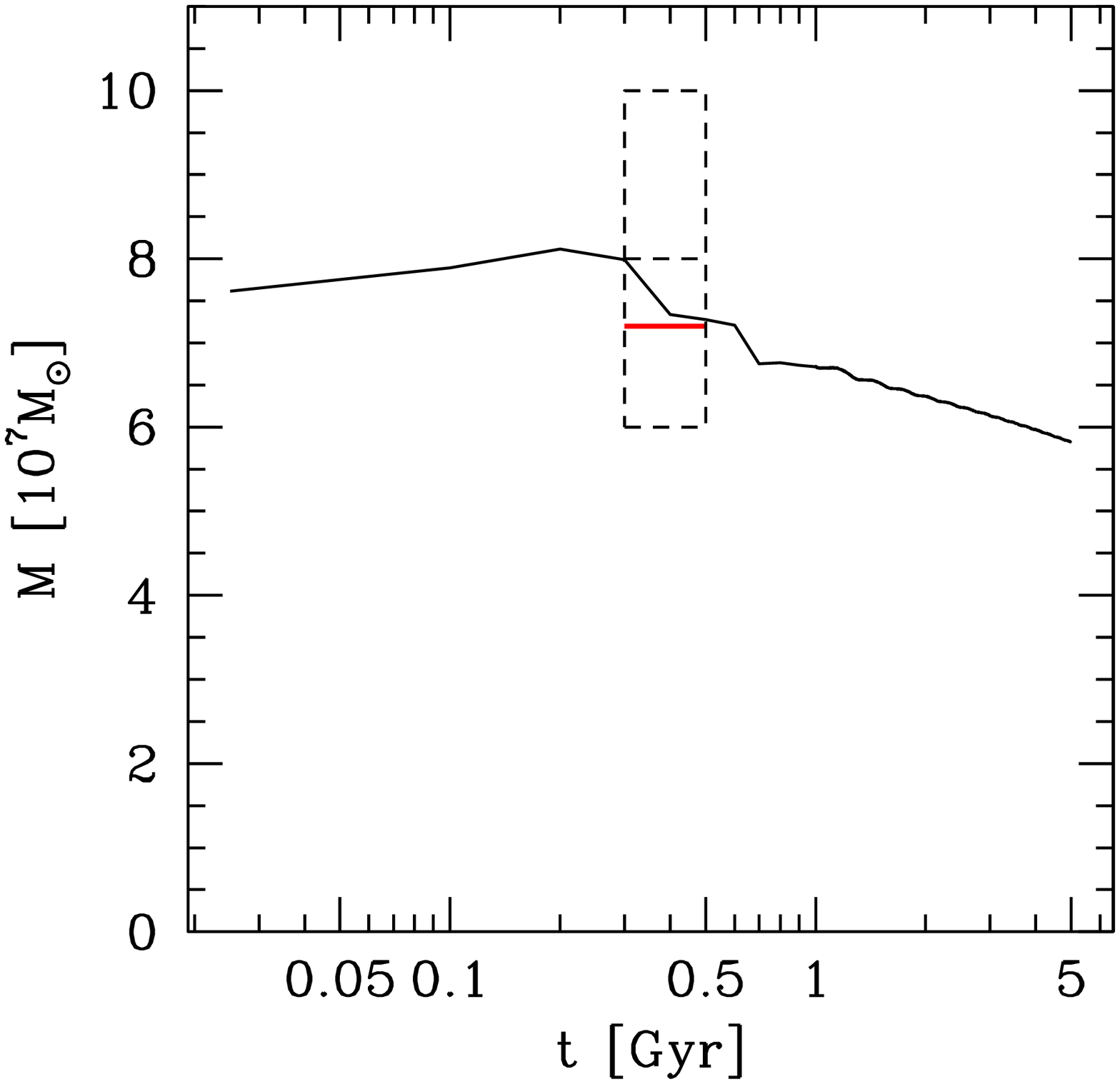}
  \epsfxsize=03.2cm
  \epsfysize=03.2cm
  \epsffile{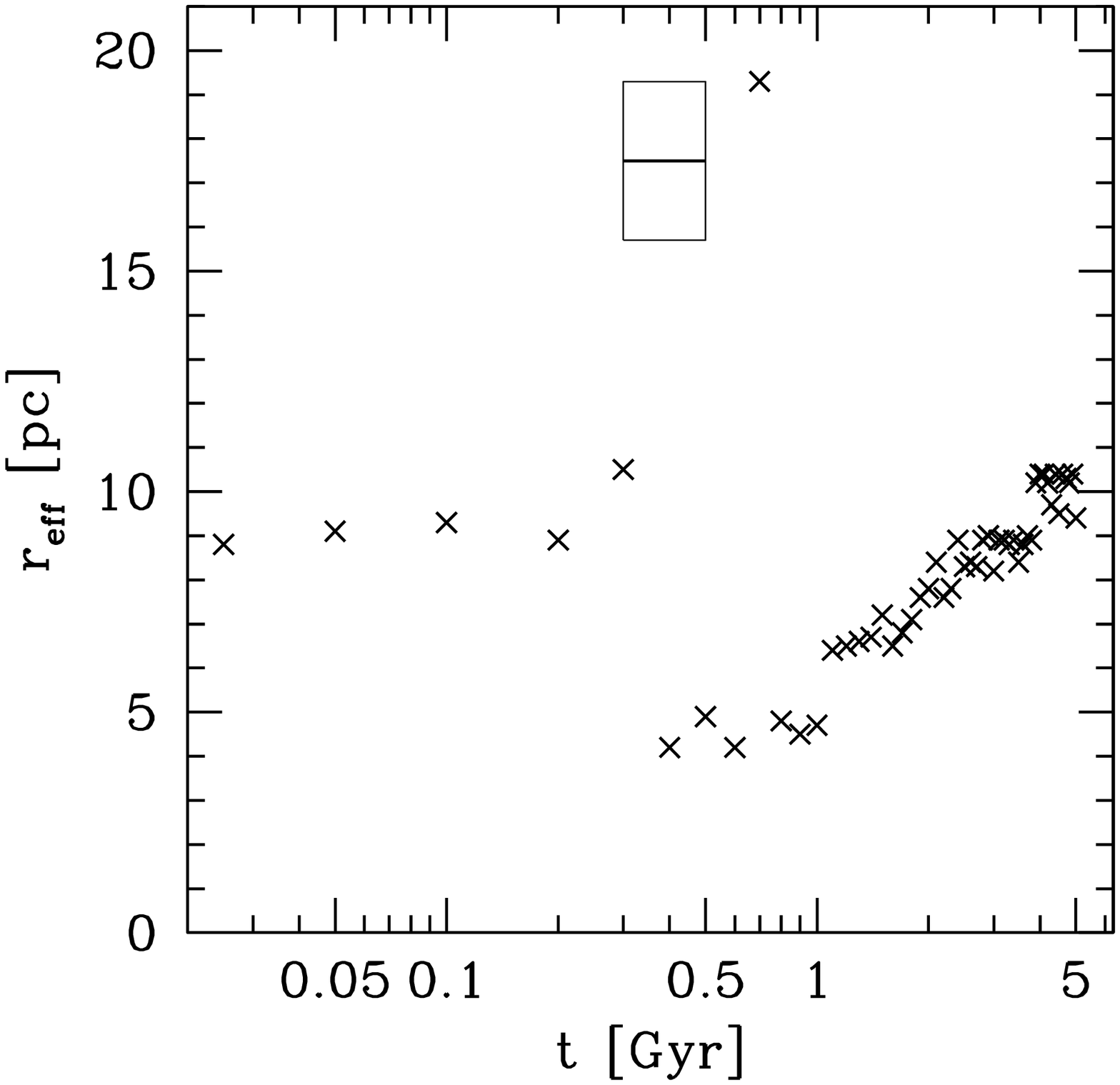}
  \epsfxsize=03.2cm
  \epsfysize=03.2cm
  \epsffile{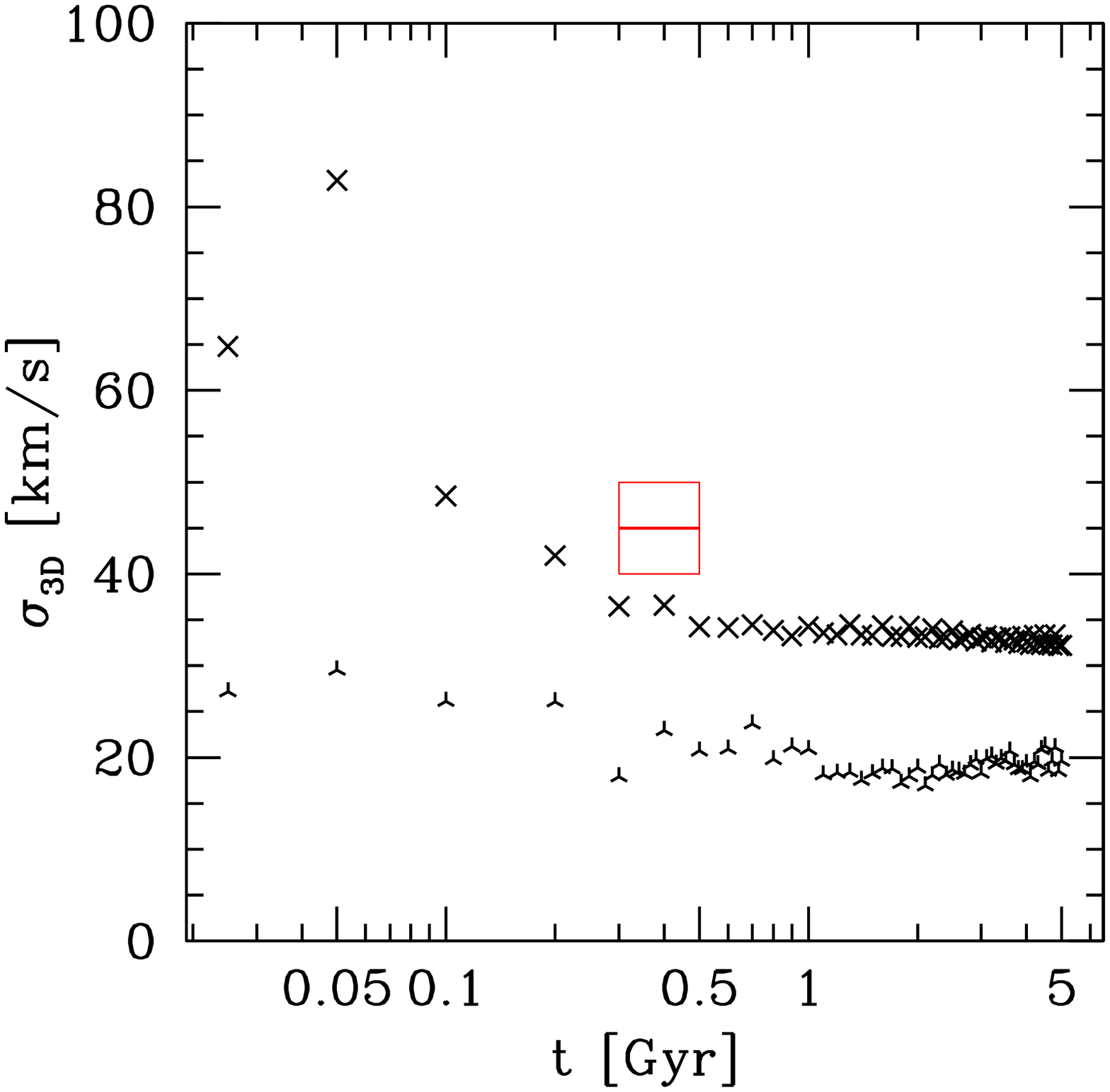}
  \caption{Time evolution of characteristic parameters of our model.
    Total mass (left panel), effective radius (middle panel), and
    velocity dispersion (right panel): Crosses are the maximum
    velocity dispersion values and three-pointed stars are the central
    values.  Horizontal lines and boxes show the measured values
    for W3 with $1\sigma$-uncertainty.} 
  \label{fig:evol}
 \vspace*{-0.5cm}
\end{figure}

In Fig.~\ref{fig:evol} we show the evolution of the total mass, of the 
effective radius and of the velocity dispersion of our model, whose
dynamical evolution was followed for $5$~Gyr.  Shown in the right
panel of the figure are the central values of the dynamically cold
core as well as the maximum values of the velocity dispersion.  The
boxes show the values measured for W3. 
The total mass as well as the maximum velocity dispersion of our model
is in reasonable agreement with the data of W3.  Only the effective
radius of our model is too small compared with W3.  It starts off at a
value of about $10$~pc and it increases slightly until the majority of
the star clusters have merged, deforming the core progressively.
After the merging process has ceased the core of the merger object
stabilises and the effective radius drops to about $5$~pc.  But note
the extremely large value at $t=700$~Myr.  At exactly that time a late
merger event of one of the remaining star clusters happened.  If our
formation scenario is correct then this points to the possibility that
we might be observing W3 at the time when a star cluster merges with
the core of the object mimicking a large effective radius.  

The further dynamical evolution is mainly governed by the tidal
shaping of the object due to its eccentric orbit.  It leads to a
successive mass loss and an increase in the effective radius as well
as a decrease in the maximum velocity dispersion, while the central
value increases slightly. 
Looking at the merger object after $5$~Gyr of its dynamical evolution
(Fig.~\ref{fig:t300}) shows that this object is very stable against
tidal disruption.  Also the dynamically cold core is not a transient
feature but survives the evolution.  The total mass of the object is
still of the order of $6 \cdot 10^{7}$~M$_{\odot}$ which is an order
of magnitude more than the most massive globular cluster
($\omega$~Cen) of the Milky Way.  The relaxation time-scale of this
object, adopting the formula from Spitzer \& Hart (1971), is
$t_{\rm relax}\approx 4000$~Gyr.  This shows that this object might to be 
thought of as a dwarf galaxy rather than a globular cluster, since
$t_{\rm relax}$ is much longer than a Hubble time.
 
\section{Discussion}

With our numerical model we introduce a formation scenario for the
ultra-massive 'star cluster' W3 found in NGC~7252.  We propose that
this object may not have formed as one star cluster, but rather as a
star cluster complex.  The star clusters in this complex have merged
and formed W3.  These star cluster complexes (super-clusters) are a
common feature in interacting late-type galaxies.

We predict how W3 should look like if it would be possible to resolve
it.  Its age places it at about the transition time between having
many remaining star clusters visible inside and the time when all star 
clusters have already merged.  One should see the last surviving star
clusters in the stage of merging.

Furthermore a merger object as massive as W3 should show a distinct
core--envelope structure.  The merger object in our calculation had a
dense, dynamically cold core and an extended envelope.  This
core--envelope structure is not a transient feature but a stable
configuration.  We followed the evolution of our merger object for
$5$~Gyr and this structure remained.  The core is formed out of the
merged star clusters.  The envelope consists out of the stripped
stars, lost by the star clusters during the initial violent merger
process.  In our previous models the merger objects are not massive
enough to keep stripped stars bound and they are lost.  In the
case of the W3 model the core object is massive enough to retain the
stripped stars bound, forming the envelope structure of our model.

W3 is said to be one of the newly discovered ultra-compact dwarf (UCD) 
galaxies and also one of the most massive ones (Maraston et al.\
2004).  Other UCDs are found and studied around the central galaxy in
the Fornax cluster (NGC~1399) by Hilker et al.\ (1999), Phillipps et
al.\ (2001) and Mieske et al.\ (2004).  There these objects are quite
old.  There are two competing formation scenarios for UCDs.  While
Bekki et al.\ (2003) propose that these objects are the remaining
cores of stripped nucleated dwarf galaxies, Fellhauer \& Kroupa
(2002a) proposed the merging scenario from star cluster complexes as a
possible formation process, resulting naturally from the merging of
gas-rich galaxies in groups as the groups merge with a galaxy cluster.
A possible way to distinguish between the two scenarios would be an
analysis of the stellar populations in these objects.  While the
merging star cluster scenario implies that the stars of the UCDs have
more or less the same metallicity and age (this holds at least for the
most prominent population formed in the starburst), cores of dwarf
galaxies should show a more complex metallicity and age distribution.
And as dwarf galaxies are believed to be the oldest building blocks in
the universe, the cores should show a prominent very old population.

In the case of W3 the stripping scenario can be completely ruled out.
Age estimates of this object range from $300$ to $500$~Myr, which is
also the age of the interaction (NGC~7252 is a merger remnant of two
gas-rich disc galaxies).  This time is much too short for a nucleus to
be stripped of its surrounding dwarf galaxy.  If this implies that all
UCDs must have formed in the same way or if both scenarios are
possible and are realized by nature can not yet be asserted.

\end{document}